\documentclass[10pt, twocolumn, comsoc]{IEEEtran}

\def\argmax{\mathop{\rm arg\,max}}
\def\argmin{\mathop{\rm arg\,min}}
\usepackage{graphicx,epsfig}
\usepackage[noadjust]{cite}
\usepackage{mcite}
\usepackage{booktabs}
\usepackage{amsfonts,helvet}
\usepackage{fancyhdr}
\usepackage{threeparttable}
\usepackage{epsf,epsfig}
\usepackage{amsthm}
\usepackage{amsmath}
\usepackage{siunitx}
\usepackage{amssymb}
\usepackage{dsfont}
\usepackage{subfigure}
\usepackage{color}
\usepackage[linesnumbered,ruled,noend]{algorithm2e}
\usepackage{algpseudocode}
\usepackage{algcompatible}
\usepackage{enumerate}
\usepackage{gensymb}
\usepackage{cancel}
\usepackage{graphicx}
\usepackage{wrapfig}
\usepackage{ragged2e}

\newtheorem{theorem}{Theorem}

\newtheorem{proposition}{Proposition}
\newtheorem{remark}{Remark}

\usepackage{bbm}
\usepackage{eucal}

\usepackage{dsfont}
\usepackage{boldline}

\def\ba{{\bf a}}

\def\bff{{\bf f}}
\def\bg{{\bf g}}
\def\bh{{\bf h}}

\def\bs{{\bf s}}

\def\bu{{\bf u}}

\def\bw{{\bf w}}
\def\bx{{\bf x}}

\def\bA{{\bf A}}

\def\bF{{\bf F}}

\def\bH{{\bf H}}

\def\bR{{\bf R}}

\def\bU{{\bf U}}
\def\bV{{\bf V}}
\def\bW{{\bf W}}

\def\cC{\mbox{$\mathcal{C}$}}

\def\cN{\mbox{$\mathcal{N}$}}

\def\cS{\mbox{$\mathcal{S}$}}

\def\bbC{\mbox{$\mathbb{C}$}}

\def\bbE{\mbox{$\mathbb{E}$}}

\begin{document}

\title{Hybrid Precoding Revisited:
Low-Dimensional Subspace Perspective for MU-MIMO Systems
}
\author{Mintaek~Oh and Jinseok~Choi

\thanks{Mintaek Oh and Jinseok Choi are with the School of Electrical Engineering, Korea Advanced Institute of Science and Technology (KAIST), Daejeon, 34141, South Korea (e-mail: {\texttt{\{ohmin, jinseok\}@kaist.ac.kr}}). 
}
}

\maketitle \setcounter{page}{1} 

\begin{abstract}
This letter presents a low-complexity hybrid precoding framework for multiuser multiple-input multiple-output (MIMO) systems by leveraging a low-dimensional subspace property.
Under the low-dimensional subspace perspective, we first identify an unconstrained optimal radio-frequency (RF) precoder. 
We then optimize a hybrid precoder via a reduced-complexity precoding method.
We further extend the proposed framework to $(i)$ a dynamic-subarray antenna partitioning algorithm that adaptively allocates subsets of antennas associated with RF chains, and $(ii)$ a channel covariance-based approach to exploit statistical channel state information at a transmitter (CSIT), ensuring robustness with partial CSIT.
Simulations validate that our proposed algorithms achieve superior performance while significantly reducing complexity compared to existing methods.
\end{abstract}
\vspace{-0.4cm}
\vspace{-0.5cm}
\section{Introduction}
\vspace{-0.1cm}
Massive multiple-input multiple-output (MIMO) systems have emerged as a cornerstone technology for next-generation wireless communications, offering substantial gains in spectral efficiency (SE) and coverage through the deployment of large antenna arrays at a base station (BS) \cite{huo2023technology}. 
In theory, simply scaling the number of BS antennas yields a near-linear SE increase under ideal conditions.
In practice, however, fully-digital precoding over  enormous antennas is often impractical because every element demands its own radio-frequency (RF) chain, driving up hardware cost and power consumption \cite{choi2020advanced}.

To address these hardware constraints, hybrid precoding architectures, which combine analog and digital processing, have been proposed as a cost-effective solution.
By employing a limited number of RF chains connected to a larger antenna array through analog phase shifters, hybrid architectures can approximate the performance of fully-digital systems while significantly reducing hardware complexity \cite{sohrabi2016hybrid}.  
Despite their promise, designing efficient hybrid precoders remains challenging due to the non-convex unit-modulus constraints imposed by analog phase shifters as well as the coupling between analog and digital precoding optimization \cite{deng2025csi}.

Conventional hybrid precoding approaches predominantly employ orthogonal matching pursuit (OMP) to construct RF precoders based on propagation path angles \cite{el2014spatially, lee2014hybrid}.
While the OMP-based hybrid precoder design simplifies the problem by casting it as a sparsity-constrained matrix reconstruction task, it is inherently limited to sparse channel environments.
{In point-to-point MIMO systems, the capacity-achieving precoder aligns with the directions of the channel eigenvectors. 
Consequently, hybrid designs seek analog and digital components that approximate this fully-digital optimum, often via alternating optimization \cite{ni2017near}. 
By contrast, in multiuser (MU)-MIMO systems, no closed-form expression for an optimal precoder is known.}
Nevertheless, it was revealed that the RF precoder determined by the phase angles of the channel matrix can attain near-optimal SE performance using a zero-forcing (ZF) precoder \cite{liang2014low} in MU-MIMO systems.
Additionally, in \cite{sohrabi2016hybrid}, a high-performance RF precoding design was proposed based on an iterative RF precoder design with a ZF digital precoder. 
However, most hybrid precoding designs rely on either linear precoders or iterative RF precoding designs.
Therefore, substantial gains remain achievable through advanced nonlinear techniques and computationally efficient approaches.

Unlike most prior works that focus on fully-connected architectures where each RF chain connects to all antennas, hybrid precoding studies have emphasized the potential benefits of partially-connected structures, which reduce hardware complexity \cite{li2019hybrid, gadiel2021dynamic, hu2023joint}. 
To deal with this, dynamic-subarray designs have been introduced, where each subarray is dynamically connected to a specific set of BS antennas  \cite{park2017dynamic, gadiel2021dynamic}.
However, these approaches typically rely on computationally intensive iterative algorithms that alternately optimize the selection of BS antennas associated with RF precoders as well as analog and digital components, making them impractical for real-time implementation. 
Moreover, a practical limitation is the common assumption of instantaneous channel state information at the transmitter (CSIT).

To address the aforementioned issues, we propose a low-complexity hybrid precoding framework based on a low-dimensional subspace perspective for multiuser MIMO (MU-MIMO) systems. 
Our key contributions are as follows:
\begin{itemize}
    \item
    By identifying a near-optimal RF precoder through a low-dimensional subspace property, we develop a low-complexity hybrid precoding framework.

    \item 
    Based on our framework, we introduce dynamic-subarray architectures that dynamically partition BS antennas into subsets associated with RF chains.

    \item 
    We further extend our RF precoder design to operate with statistical CSIT, enabling robust performance.
\end{itemize}

Our proposed hybrid precoding methods are evaluated through simulations. 
The results demonstrate that our methods achieve significant complexity reduction while maintaining SE performance compared to state-of-the-art methods, for both fully-connected and dynamic-subarray architectures.


{\textit{Notation}}:
${\bf{I}}_N$ is the identity matrix of size $N \times N$ and $\bf 0$ is a zero vector with proper dimension.
$\|\bA\|_F$ represents Frobenius norm.
We use ${\rm{tr}}(\cdot)$ for trace operator, ${\rm{rank}}(\cdot)$ for matrix rank, and ${\rm arg}(\cdot)$ for the argument of a complex number.
With mean $m$ and variance $\sigma^{2}$, we use $\cC\cN(m,\sigma^{2})$ for a circularly symmetric complex Gaussian distribution.

\section{System Model and Problem Formulation} \label{sec:system}

\subsection{System Model}
\vspace{-0.05cm}
We consider a narrowband downlink massive MIMO system with $N$ BS antennas, $N_{\sf RF}$ RF chains, and $K$ single-antenna users where $N > N_{\sf RF}$.
The hybrid precoding architecture operates in two stages. 
First, the digital baseband precoder $\bF_{\sf D}=[\bff_{\sf D,1}, \cdots, \bff_{\sf D,K}] \in \bbC^{N_{\sf RF} \times K}$ processes $K$ data streams.
These digitally precoded signals are then upconverted through $N_{\sf RF}$ RF chains and fed to an analog RF precoder $\bF_{\sf RF} \in \bbC^{N \times N_{\sf RF}}$, which is implemented using phase shifters satisfying $|\bF_{\sf RF}(i,j)|=1$ for all elements.
In this regard, the overall hybrid precoder is $\bF = \bF_{\sf RF}\bF_{\sf D} \in \bbC^{N \times K}$.
The transmitted signal vector is expressed as $\bx = \bF \bs$ where $\bs = [s_1, s_2, \ldots, s_K]^{\sf T}$ contains the data symbols with $s_k \sim \mathcal{CN}(0,P),\; \forall k$.

Assuming all channels have the same number of propagation paths $L$, we adopt a geometric channel model as
\begin{align}
    \label{eq:channel}
    \bh_k = \frac{1}{\sqrt{L}} \sum_{\ell = 1}^{L} \alpha_{k,\ell} \ba(\theta_{k,\ell}), \forall k,
\end{align}
where $\theta_{k,\ell}\in[-\pi,\pi]$ is the $\ell$th path's angle of departure (AoD), $\ba(\cdot)$ is an array response vector at the BS, and {\color{black}$\alpha_{k,\ell} \sim \cC \cN(0, 1)$} is a complex path gain.
The overall channel matrix is $\bH = [\bh_1,\bh_2, \cdots, \bh_K]\in \bbC^{N \times K}$.
We consider uniform linear arrays at the BS in which the array response vector is defined as $\ba(\theta) = \left[1, e^{j\frac{2\pi d}{\lambda}\sin(\theta)},\ldots, e^{j(N - 1)\frac{2\pi d}{\lambda}\sin(\theta)} \right]^{\sf T}$, 
where $\lambda$ is a signal wavelength and $d$ is a distance between antenna elements.
The received signal of user $k$ is given by
\begin{align}
    \label{eq:y_k}
    y_k = \bh_k^{\sf H} \left(\bF_{\sf RF} \bff_{{\sf D},k} \right)s_k + \sum_{\ell=1,\ell \neq k}^{K} \bh_k^{\sf H}\left(\bF_{\sf RF} \bff_{{\sf D},\ell} \right)s_{\ell} + n_k,
\end{align}
where $n_k \sim \cC \cN(0,\sigma^2)$ is additive white Gaussian noise.
We assume the perfect CSIT scenario unless mentioned otherwise.

\subsection{Problem Formulation}
\vspace{-0.1cm}

Based on \eqref{eq:y_k}, the SE of user $k$ is given by
\begin{align}
    R_k = \log_2 \left(1+ \frac{\left|\bh_k^{\sf H}\left(\bF_{\sf RF} \bff_{{\sf D},k} \right)\right|^2}{\sum_{\ell=1,\ell \neq k}^{K}\left|\bh_k^{\sf H}\left(\bF_{\sf RF} \bff_{{\sf D},\ell} \right)\right|^2 + \sigma^2/P} \right).
\end{align}
We formulate the SE maximization problem as
\begin{align} 
    \label{eq:main_problem}
    \mathop{{\text{maximize}}}_{{\bF_{\sf D},\bF_{\sf RF}}}& \;\;  \sum_{k=1}^{K} R_k
    \\
    \label{eq:power_constraint}
    {\text{subject to}} & \;\; {\rm tr}\left(\bF_{\sf RF} \bF_{\sf D} \bF_{\sf D}^{\sf H} \bF_{\sf RF}^{\sf H}\right) = 1,
    \\
    \label{eq:unitmodulus_constraint}
    &\; \left|\bF_{\sf RF}(i,j)\right| = 1,\;\; \forall i,j.
\end{align}
The problem in \eqref{eq:main_problem} is inherently non-convex with the non-convex unit-modulus constraint in \eqref{eq:unitmodulus_constraint}, making it  intractable to solve optimally.

\section{Fully-connected Hybrid Architectures}
\vspace{-0.05cm}
This work presents that hybrid precoding architectures can implement fully-digital precoding with potentially fewer RF chains. 
To achieve the fully-digital precoder, the number of RF chains must satisfy $N_{\sf RF} \geq K$ \cite{sohrabi2016hybrid}.
This fundamental constraint means that hybrid architectures must have \emph{at least} as many RF chains as user streams to replicate fully-digital performance.
From a hardware‑efficiency standpoint, provisioning \(\!N_{\sf RF}>K\) would only increase cost and power consumption without yielding additional beamforming gains.  
Thus, we set \(N_{\sf RF}=K\), i.e., the minimum number of RF chains capable of realizing the fully-digital precoder $\bF_{\sf opt}$ unless mentioned otherwise.

\subsection{Optimizing RF Precoder $\bF_{\sf RF}$} \label{subsec:RF_precoder}
\vspace{-0.1cm}
Throughout this letter, our goal is to find the precoders that are most similar to the fully-digital precoder $\bF_{\sf opt}$ in the decomposed form, i.e., $\bF_{\sf opt} \approx \bF_{\sf RF} \bF_{\sf D}$.
To this end, we introduce the following low-dimensional subspace property of the optimal fully-digital precoder.
\begin{theorem}
    \label{th:HW}
    (Low-dimensional subspace property \cite{zhao2023rethinking})
    Any nontrivial stationary point with the optimal precoder in \eqref{eq:main_problem} must lie in the range space of $\bH$: $\bF_{\sf opt}^{\star} \!=\! \bH \bW^{\star}$ with $\bW^{\star}\!\!\in \bbC^{K \!\times\! K}$.
\end{theorem}
Based on Theorem~\ref{th:HW}, the precoders in the hybrid structures can be reorganized to yield the optimal RF precoder, which is presented in the following proposition.
\begin{proposition}
    The near-optimal RF precoder $\bF_{\sf RF}^{\star}$ in \eqref{eq:main_problem} is achieved by $e^{j {\rm arg}(\bU)}$ where $\bU$ is the left-singular vectors of $\bH$.
    \begin{proof}
    Based on Theorem~\ref{th:HW}, in hybrid precoding architectures, the optimal fully-digital precoder is expressed as 
    \begin{align}
        \label{eq:HW}
        \bF_{\sf opt}^{\star} = \bH \bW^{\star} \approx \bF_{\sf RF} \bF_{\sf D}
        = \bF_{\sf RF} \Tilde{\bH} \Tilde{\bW} = \bF_{\sf RF} \bF^{\sf H}_{\sf RF}\bH \Tilde{\bW},
    \end{align}
    where $\Tilde{\bH} = \bF^{\sf H}_{\sf RF}\bH$ is the effective channel for the digital baseband precoder and the optimal  digital baseband precoder is given by $\bF_{\sf D} = \Tilde{\bH} \Tilde{\bW}$.
    Using singular value decomposition (SVD) of the channel matrix: $\bH = \bU \boldsymbol{\Sigma}\bV^{\sf H}$, from \eqref{eq:HW}, we have
    \begin{align}
        \label{eq:opt_precoder}
        \bF_{\sf opt}^{\star} = \bU \boldsymbol{\Sigma}\bV^{\sf H} \bW^{\star} \approx \bF_{\sf RF} \bF^{\sf H}_{\sf RF} \bU \boldsymbol{\Sigma}\bV^{\sf H} \Tilde{\bW}.
    \end{align}
    In \eqref{eq:opt_precoder}, it is obvious that the optimal RF precoder is the left-singular matrix $\bU \in \bbC^{N \times K}$ because of ${\rm rank}(\bH) = K$.
    Therefore, the RF precoder optimization problem is given by
    \begin{align}
        \label{eq:RF_problem}
        \mathop{{\text{minimize}}}_{{\bF_{\sf RF}}}& \;\;  \left\|\bF_{\sf RF} - \bU\right\|_F^2
        \\
        \label{eq:RF_constraint}
        {\text{subject to}} & \;\; \left|\bF_{\sf RF}(i,j)\right| = 1,\forall i,j.
    \end{align}
    To solve \eqref{eq:RF_problem}, we consider element-wise optimization as
    \begin{align}
        &\left|\bF_{\sf RF}(i,j) - \bU(i,j)\right|^2 =  
        \label{eq:RF_problem_2} \\ \nonumber
        &\quad 1 \!+\! |\bU(i,j)|^2 \!-\! 2|\bU(i,j)|\cos ({\rm arg}(\bF_{\sf RF}(i,j)) \!-\! {\rm arg}(\bU(i,j))).
    \end{align}
    The optimal solution of \eqref{eq:RF_problem_2} is straightforward: ${\rm arg}(\bF_{\sf RF}(i,j)) = {\rm arg}(\bU(i,j)),\forall i, j$.
    Therefore, the optimal RF precoder with the unit-modulus constraint  \eqref{eq:RF_constraint} is achieved by extracting phase angles of $\bU$: $\bF_{\sf RF}^{\star} = e^{j {\rm arg}(\bU)}$.
    \end{proof}
\end{proposition}

\subsection{Optimizing Digital Baseband Precoder $\bF_{\sf D}$}
\vspace{-0.1cm}
\begin{algorithm}[t]
    \caption{R-WMMSE-based Digital Precoding}
    \label{alg:algorithm_1} 
    {\bf{initialize}}: $\Tilde{\bW}^{(0)}$.
    \\
    Set the iteration count $t= 1$.
    \\
    \While {$\left\| \Tilde{\bW}^{(t)} \!-\! \Tilde{\bW}^{(t-1)} \right\|_F /\|\Tilde{\bW}^{(t-1)}\|_F > \varepsilon$ {\rm \&} $t \leq t_{\max} $}{
    Compute $\bu$ and $\pmb{\lambda}$ by \eqref{eq:RWMMSE_u} and  \eqref{eq:RWMMSE_lamb}, respectively.
    \\
    Update $\Tilde{\bW}^{(t)}$ by \eqref{eq:RWMMSE_w}.
    \\
    $t \leftarrow t+1$.}
    $\Tilde{\bW}^{\star} \leftarrow \Tilde{\bW}^{(t)}$.
    \\
    \Return{\ }{${\bF}_{\sf D}^{\star} \leftarrow \Tilde{\bH}\Tilde{\bW}^{\star}$}.
\end{algorithm}
To optimize the baseband digital precoder, we adopt the reduced-complexity weighted minimum mean square error (WMMSE) method, called R-WMMSE \cite{zhao2023rethinking}.
This method optimizes $\Tilde{\bW}$ such that $\bF_{\sf D} = \Tilde{\bH}\Tilde{\bW}$ leveraging the low-dimensional subspace property.
The sum SE maximization problem can be reduced with given $\bF_{\sf RF}$ to the following sum mean square error (MSE) minimization problem \cite{zhao2023rethinking}:
\begin{align} 
    \label{eq:MSE_problem}
    \mathop{{\rm minimize}}_{\bu, \pmb{\lambda}, \Tilde{\bW}}& \;\;  \sum_{k=1}^{K}    \left(\lambda_k e_k - \log \lambda_k\right),
\end{align}
where $\bu = [u_1, u_2, \ldots, u_K]^{\sf T}$ and $\pmb{\lambda} = [\lambda_1, \lambda_2, \ldots, \lambda_K]^{\sf T}$ are auxiliary variables, and $e_k = (1 - u_k^{*}\bh_k^{\sf H}\bH \Tilde{\bw}_k)^2 + |u_k|^2\left(\sum_{j\neq k}\bh_k^{\sf H}\bH \Tilde{\bw}_j \Tilde{\bw}_j^{\sf H}\bh_k \bH^{\sf H} + \frac{\sigma^2}{P} \sum_{i=1}^{K} {\rm tr}\left(\bH^{\sf H}\bH \Tilde{\bw}_k \Tilde{\bw}_k^{\sf H}\right)\right)$ with $\Tilde{\bW} = \left[\Tilde{\bw}_1, \Tilde{\bw}_2, \ldots, \Tilde{\bw}_K\right]$.
The objective function of \eqref{eq:MSE_problem} is convex in each variable $(\bu, \bw, \Tilde{\bW})$, respectively.
Hence, one variable at each iteration is solved while fixing the others; thereby the subproblem with respect to each variable can be globally solved.
According to \cite{zhao2023rethinking}, the problem in \eqref{eq:MSE_problem} is solved by updating $\bu, \pmb{\lambda}$, and $\Tilde{\bW}$ with the following closed-form at each iteration:
\begin{align}
    \label{eq:RWMMSE_u}
    &u_k = \left(\sum_{i=1}^K\frac{\sigma^2}{P}\bH^{\sf H}\bH \Tilde{\bw}_i\Tilde{\bw}_i^{\sf H} \!+\! \sum_{j=1}^K\bg_{k}^{\sf H} \Tilde{\bw}_j \Tilde{\bw}_j^{\sf H}\bg_{k} \right)^{-1}\bg_{k}^{\sf H} \Tilde{\bw}_k, \forall k,
    \\
    \label{eq:RWMMSE_lamb}
    &\lambda_k = \left(1 - u_k^{*}\bg_{k}^{\sf H} \Tilde{\bw}_k \right)^{-1} , \forall k,
    \\
    \label{eq:RWMMSE_w}
    &\Tilde{\bw}_k \!=\! \left(\!\sum_{i=1}^K\! \frac{\sigma^2}{P}|u_i|^2 \lambda_i \bH^{\sf H}\bH \!+\! \sum_{j=1}^K\! |u_j|^2 \lambda_j \bg_{j} \bg_{j}^{\sf H} \!\right)^{-1}\!\!\!\!\!\bg_{k} u_k \lambda_k, \forall k\!,
\end{align}
where $\bg_{k} =\bH^{\sf H}\bh_k$.
We repeat these steps until either converges or reaches the maximum iteration count $t_{\rm max}$.
Based on an initial precoder $\bF_{\sf D}^{(0)} = \Tilde{\bH}\Tilde{\bW}^{(0)}$, we use $\Tilde{\bW}^{(0)}$ as an initial matrix.
We describe the R-WMMSE algorithm in Algorithm~\ref{alg:algorithm_1}.

\subsection{Complexity Analysis} \label{subsec:complexity}
\vspace{-0.1cm}
We summarize our proposed hybrid precoding algorithm in Algorithm~\ref{alg:algorithm_2}.
The algorithm first extracts phase angles from $\bU$ to construct $\bF_{\sf RF}$, requiring  $\CMcal{O}(N K^2)$ complexity via SVD.
Subsequently, the digital precoder $\bF_{\sf D}$ is updated using R-WMMSE, where R-WMMSE requires $\CMcal{O}(T(K^3 + N_{\sf RF}K^2)$ complexity with $T$ iterations \cite{zhao2023rethinking}.
Therefore, the overall complexity of the proposed algorithm is $\CMcal{O}(NK^2 + T(K^3 + N_{\sf RF}K^2))$, which scales linearly with respect to $N$.
This represents a significant improvement over existing approaches \cite{li2019hybrid, hu2023joint, ma2025digital}, which typically exhibit $\CMcal{O}(N^2)$ or $\CMcal{O}(N^3)$ complexity.

\begin{algorithm}[t] 
    \label{alg:algorithm_2}
    \caption{Proposed Hybrid Precoding}
    {\bf{initialize}}: $\bF_{\sf D}^{(0)}$.
    \\
    Compute $\bH = \bU \boldsymbol{\Sigma}\bV^{\sf H}$.
    \\
    Set $\bF_{\sf RF} = e^{j {\rm arg}(\bU)}$.
    \\
    Update ${\bF}_{\sf D} \leftarrow$ Algorithm~\ref{alg:algorithm_1}$(\Tilde{\bH})$ with $\Tilde{\bH} = \bF_{\sf RF}^{\sf H}\bH$.
    \\
    Normalize $\bF_{\sf D} = \frac{{\bF}_{\sf D}}{\sqrt{{\rm tr}\left({\bF}_{\sf RF}{\bF}_{\sf D} {\bF}_{\sf D}^{\sf H} \bF_{\sf RF}^{\sf H}\right)}}$.
    \\
    \Return{\ }{$\bF_{\sf RF}^{\star} \leftarrow \bF_{\sf RF}$ and $\bF_{\sf D}^{\star} \leftarrow \bF_{\sf D}$}.
\end{algorithm}
\section{Extension to Practical Hybrid Architectures} \label{sec:extension}
\vspace{-0.05cm}

\subsection{Dynamic-Subarray Hybrid Precoding} \label{subsec:dynamic}
\vspace{-0.1cm}
In this subsection, we propose the hybrid precoding strategy for partially-connected architectures, where each RF chain connects to only a subset of BS antennas.
We assume that $N$ is divisible by $N_{\sf RF}$, with all RF chains connected to equally-sized subsets of $N_{\sf sub} = \frac{N}{N_{\sf RF}}$ antennas, thereby preserving the unitary property of $\bF_{\sf RF}$. 
Let $\cS_r$  represent the subset of antenna indices connected to the $r$th RF chain.
To partition such subset, the straightforward approach is to sequentially assign the fixed antenna indices: $\cS_1 = \{1, \cdots, N_{\sf sub}\}, \cdots, \cS_{N_{\sf RF}} = \{(N_{\sf RF}-1)N_{\sf sub} + 1, \cdots, N_{\sf RF}N_{\sf sub}\}$.
With this fixed-subarray architecture, the RF precoder becomes block-diagonal as
\begin{align}
    \bF_{\sf RF} = {\rm blkdiag}\left(\bff_{{\sf RF}, \cS_1}, \bff_{{\sf RF}, \cS_2}, \ldots, \bff_{{\sf RF}, \cS_{N_{\sf RF}}}\right),
\end{align}
where $\bff_{{\sf RF}, \cS_{r}}$ is a RF precoding vector associated with the $r$th RF chain.
However, this fixed-subarray structure may not be optimal because the antenna assignment is predetermined regardless of channel conditions. 
To overcome this limitation, we formulate a dynamic-subarray partitioning problem that adaptively selects subsets based on channel conditions:
\begin{align}
    \label{eq:dynamic_problem}
    &\left\{ \cS_r^{\star} \right\}_{r=1}^{N_{\sf RF}} = \argmax_{\cS_1, \ldots, \cS_{N_{\mathrm{RF}}}} \sum_{k=1}^K R_k
    \\
    \label{eq:dynamic_constraint1}
    &\text{subject to}\; \bigcup_{r=1}^{N_{\mathrm{RF}}} \cS_r = \{1, 2, \cdots, N\}, 
    \\
    \label{eq:dynamic_constraint2}
    & \qquad \qquad \; \cS_i \cap \cS_j = \emptyset \quad \text{for } i \neq j, 
    \\
    \label{eq:dynamic_constraint3}
    & \qquad \qquad \; |\cS_r| = N_{\sf sub} \quad \forall r.
\end{align}
This combinatorial optimization problem requires searching over all possible subarray configurations.
The total number of possible combinations is given by $\frac{N!}{(N_{\sf sub}!)^{N_{\sf RF}} N_{\sf RF}!}$.
For instance, when $N\!=\!32$ and $N_{\sf RF} \!=\! 4$, the number of possible configurations is approximately $4.15 \!\times\! 10^{15}$.
Therefore, an efficient approach is essential to make this problem tractable in practice.

Given this subarray architecture, our goal is to design the RF precoder $\bF_{\sf RF}$ to approximate a desired unitary matrix $\bU$ as closely as possible.
Accordingly, under the constraints in \eqref{eq:dynamic_constraint1}--\eqref{eq:dynamic_constraint3}, the optimization problem in \eqref{eq:dynamic_problem} can be reformulated as
\begin{align}
    \label{eq:dynamic_problem}
    &\left\{ \cS_r^{\star} \right\}_{r=1}^{N_{\sf RF}} = \argmin_{\cS_1, \ldots, \cS_{N_{\sf RF}}} \| \bF_{\sf RF} - \bU \|_F.
\end{align}
To solve this problem, we leverage the structure of $\bU$.
Specifically, for each RF chain $r$, we select the antenna indices corresponding to $N_{\sf sub}$ largest-magnitude elements in $r$th column of $|\bU|$.
This approach is motivated by the observation that elements of larger magnitude contribute more significantly to the precoding performance, and assigning these antennas to the corresponding RF chains helps minimize  $\| \bF_{\sf RF} - \bU \|_F$.

\begin{algorithm}[t]
    \caption{Dynamic-Subarray Partitioning}
    \label{alg:algorithm_3} 
    {\bf{initialize}}: ${\bF}_{\sf RF} = {\bf 0}$ and an available antenna set $\CMcal{A} = \{1, 2, \ldots, N\}$.
    \\
    \For {$1$ {\rm to} $N_{\sf RF}$}{
    Sort antenna indices by $|\bU(:,\!r)\!|$ in descending order\!.
    \\
    Select top $N_{\sf sub}$ antennas from $\CMcal{A}$ as $\cS_r$.
    \\
    Set $\bF_{\sf RF}(n,r) = e^{j{\rm arg}(\bU(n,r))}$ for $n \in \cS_r$.
    \\
    Update $\CMcal{A} \leftarrow \CMcal{A} \setminus \cS_r$.
    }
    {\bf{return}} $\bF_{\sf RF}$.
\end{algorithm}
We describe the proposed dynamic-subarray partitioning algorithm in Algorithm~\ref{alg:algorithm_3}.
Given $\bU$ from $\bH = \bU \boldsymbol{\Sigma} \bV^{\sf H}$ where the singular values in $\boldsymbol{\Sigma}$ are arranged in descending order, the algorithm first initializes the RF precoder and an available antenna set $\CMcal{A}$.
For each RF chain, the algorithm sorts antenna indices in descending order based on $|\bU(:, r)|$ where $\bU(:,r)$ denotes the $r$th column of $\bU$.
It then selects the top $N_{\sf sub}$ antennas with the largest magnitude values from $\CMcal{A}$ to form the subset $\cS_r$.
These selected antennas are assigned to the RF precoder by setting $\bF_{\sf RF}(n,r) = e^{j{\rm arg}(\bU(n,r))}$ for all $n \in \cS_r$.
After each iteration, the algorithm removes the selected antennas from  $\CMcal{A}$ to ensure each antenna is assigned to only one RF chain. 
This process continues until all RF chains have been allocated their antenna subsets.

This dynamic-subarray approach significantly reduces the computational complexity from $\CMcal{O}\left(\frac{N!}{(N_{\sf sub}!)^{N_{\sf RF}}N_{\sf RF}!}\right)$ for exhaustive search to $\CMcal{O}(N_{\sf RF}N\log N)$ in the proposed algorithm (Algorithm~\ref{alg:algorithm_3}).
Consequently, the overall complexity of our dynamic-subarray hybrid precoder is $\CMcal{O}(NK^2 + N_{\sf RF}N\log N + T(K^3 + N_{\sf RF}K^2))$, which avoids iterative processes in subarray partitioning while maintaining quasi-linear scaling in $N$ and linear scaling in $N_{\sf RF}$. 


\subsection{Covariance-based Approach} \label{subsec:Cov}
\vspace{-0.1cm}
In practice, it is challenging for the phase-shifter networks to re-steer their beams each channel coherence block to track small-scale fading due to its analog nature and high dimension.
A practical alternative can be a two‑stage design: $(i)$ optimize and fix the RF precoder using the slowly varying statistical CSIT (e.g., channel covariance), and $(ii)$ design the low‑dimensional digital precoder from instantaneous CSIT with this RF precoder  \cite{park2019spatial}.
Motivated by this, we extend our framework that exploits the channel covariance, thereby avoiding frequent analog re-steering while relying on instantaneous CSIT only at the digital stage.

Let $\bR = \bbE[\bH\bH^{\sf H}]$ represent the channel covariance matrix and its eigenvalue decomposition is $\bR = \bV_{\bR} \boldsymbol{\Lambda}_{\bR} \bV_{\bR}^{\sf H}$.
We now consider the channel matrix as $\bH = \bA \bH_{\sf b}$ where $\bA \in \bbC^{N \times L_{\sf tot}}$ is an array response matrix and $\bH_{\sf b} \in \bbC^{L_{\sf tot} \times K}$ is a beam domain channel matrix with $L_{\sf tot} = LK$.
The sample channel covariance matrix over a sample time $T_s$ is defined as
\begin{align}
    \Tilde{\bR} &= \frac{1}{T_s} \sum_{i=1}^{T_s} \bH[i]\bH[i]^{\sf H} = \bA \left(\frac{1}{T_s} \sum_{i=1}^{T_s} \bH_{\sf b}[i] \bH_{\sf b}[i]^{\sf H}\right) \bA^{\sf H},
\end{align}
where $\bH[i]$ and $\bH_{\sf b}[i]$ denote the channel matrices at time instant $i$.
Here, the rank of $\Tilde{\bR}$ is at most $\min (N, L_{\sf tot})$.
As high-frequency systems will employ large-antenna arrays, channels are expected to be sparse.
Consequently, the number of channel paths $L$ will likely be less than the number of antennas, i.e., $L_{\sf tot} \ll N$.
Using channel covariance information, the near-optimal RF precoder becomes $\bF_{\sf RF}^{\star} = e^{j{\rm arg}(\bV_{\bR})}$.

{
Without access to instantaneous CSIT, we can design $\bF_{\sf D}$  based on the sample average approximation (SAA) technique~\cite{joudeh2016sum} under imperfect CSIT.
In this approach, channel errors are averaged out within the WMMSE-based algorithm by exploiting the channel error covariance.
}

\begin{remark}
    \normalfont This covariance-based design plugs directly into our hybrid framework including dynamic-subarray architectures and requires $N_{\sf RF} \geq L_{\sf tot}$ RF chains instead of  $K$.
\end{remark}

\section{Numerical Results} \label{sec:numerical}
\vspace{-0.1cm}

In this section, we present numerical comparisons between our proposed algorithms and baseline algorithms:
\begin{itemize}
    \item \textbf{WMMSE (fully-digital)}: The fully-digital WMMSE-based  precoder with $N = N_{\sf RF}$.

    \item \textbf{OMP~\cite{el2014spatially}}: Leveraging WMMSE (fully-digital), the OMP-based hybrid precoder \cite{el2014spatially}.

    \item \textbf{FC~\cite{sohrabi2016hybrid}}: The fully-connected hybrid precoder based on the ZF digital precoder with $N_{\sf RF} = K+1$ \cite{sohrabi2016hybrid}.

    \item \textbf{PC~\cite{hu2023joint}}: The partially-connected hybrid precoder based on unit-modulus analog precoding \cite{hu2023joint}.

    \item \textbf{Proposed}: The proposed fully-connected hybrid precoder (Algorithm~\ref{alg:algorithm_2}).

    \item \textbf{Proposed (dynamic)}: The proposed dynamic-subarray hybrid precoder (Algorithm~\ref{alg:algorithm_3}).

    \item \textbf{Proposed (fixed)}: The proposed hybrid precoder with the fixed-subarray architectures.

    \item \textbf{Proposed-Cov}: The proposed hybrid precoder with the covariance-based approach as described in Section~\ref{subsec:Cov}.
\end{itemize}
We set $t_{{\rm max}} = 30$, $\varepsilon = 0.01$, $ d/\lambda = 0.5$, and randomly generated AoDs $\theta_{k,\ell}\in[-\pi,\pi],\forall k, \ell$.
We use ZF as an initial precoder for the WMMSE-based precoders.

\begin{figure}[t]    
    {\centerline{\resizebox{0.85\columnwidth}{!}{\includegraphics{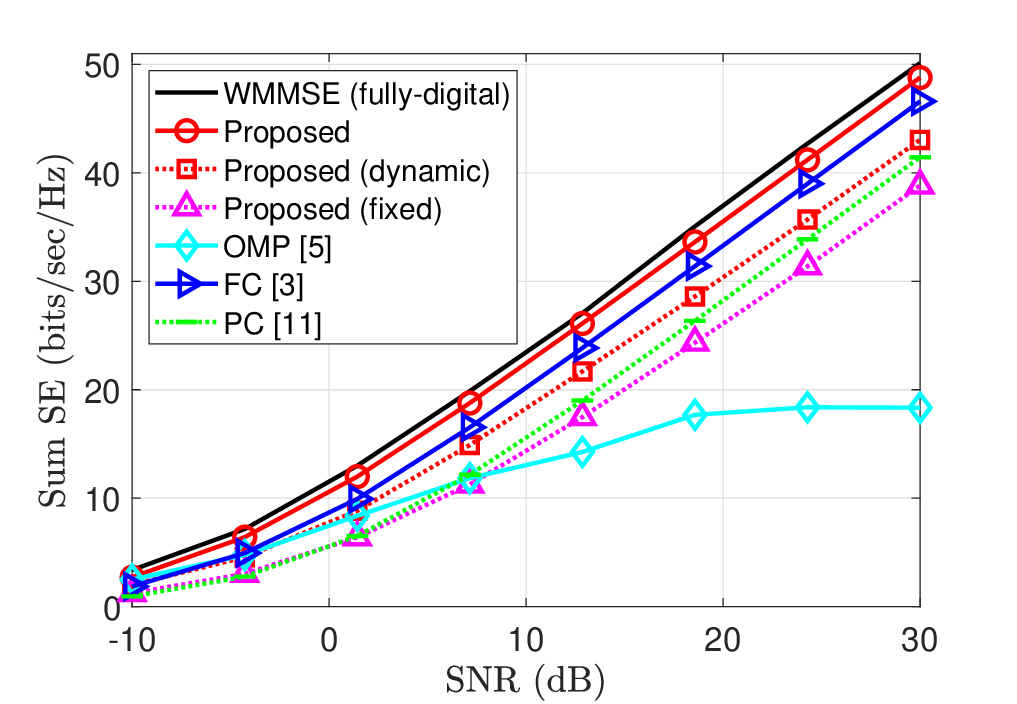}}}
    \vspace{-0.4cm}
    \caption{The sum SE versus SNR for $N = 32$ BS antennas, $K = 4$ users,  $N_{\sf RF} = 4$ RF chains, and $L = 4$ propagation paths.}
    \label{fig:PvsSE}}
\end{figure}
In Fig.~\ref{fig:PvsSE}, we consider  $N = 32, K=4, N_{\sf RF} = 4$, and $L=4$.
Fig.~\ref{fig:PvsSE} shows that the proposed fully-connected hybrid precoder achieves the near-optimal SE performance while outperforming the baseline methods.
OMP~\cite{el2014spatially} exhibits substantially degraded performance due to the mismatch between $N_{\sf RF}$ and $L_{\sf tot}$. 
Since the proposed algorithms exploit the near-optimal RF precoder design based on the low-dimensional subspace property, they achieve better SE performance than FC~\cite{sohrabi2016hybrid} in fully-connected structures and PC~\cite{hu2023joint} in subarray structures, respectively.
In particular, Proposed (dynamic) outperforms Proposed (fixed), validating the effectiveness of our dynamic-subarray strategy in subarray architectures.

\begin{figure}[t]    
    {\centerline{\resizebox{0.85\columnwidth}{!}{\includegraphics{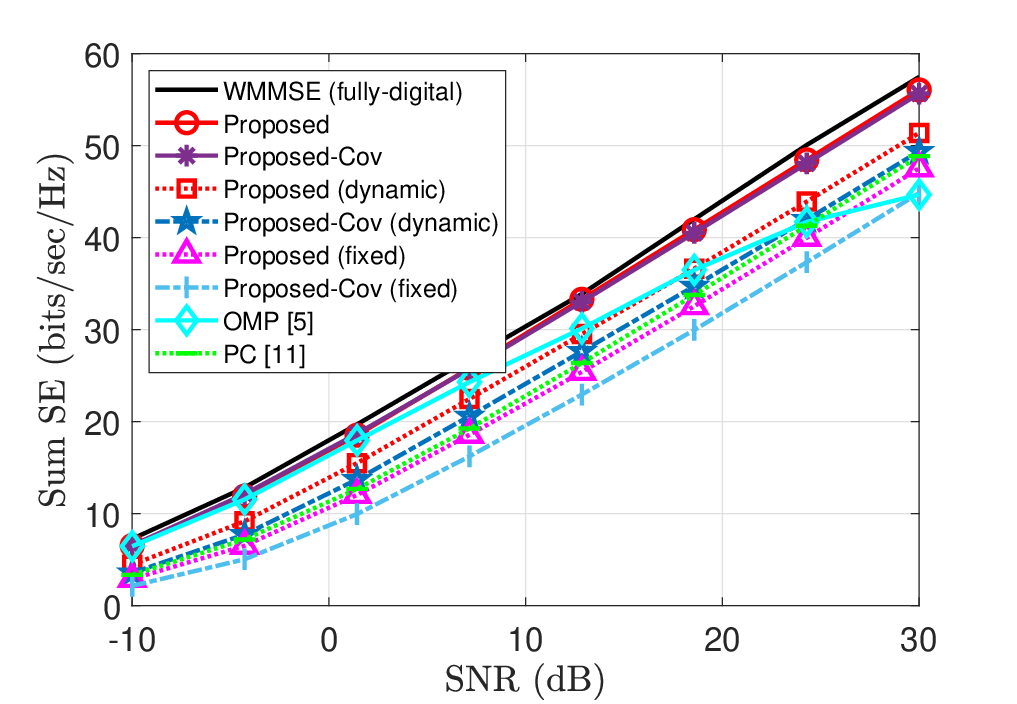}}}
    \vspace{-0.4cm}
    \caption{The sum SE versus SNR for $N = 128$ BS antennas, $K = 4$ users,  $N_{\sf RF} = 8$ RF chains, and $L = 2$ propagation paths.}
    \label{fig:PvsSE_Cov}}
\end{figure}
In Fig.~\ref{fig:PvsSE_Cov}, we consider $N = 128, K=4, L=2$, and $N_{\sf RF} = 8$.
Fig.~\ref{fig:PvsSE_Cov} shows that the proposed covariance-based algorithms demonstrate superior performance by exploiting statistical CSIT.
In particular, Proposed-Cov achieves similar SE performance to its perfect CSIT scheme because $N_{\sf RF}$ is sufficient to accommodate all propagation paths.
Similarly, OMP~\cite{el2014spatially} provides near-optimal performance in low-to-medium SNR due to adequate $N_{\sf RF}$.
However, despite having sufficient RF chains, OMP~\cite{el2014spatially} experiences significant performance degradation at high SNR. 
Our proposed dynamic-subarray algorithms consistently outperform their fixed-subarray counterparts.
Notably, Proposed-Cov (dynamic) outperforms PC~\cite{hu2023joint} even with statistical CSIT.
These results verify that the proposed algorithms maintain superior performance over conventional methods regardless of whether instantaneous or statistical CSIT is utilized, thus validating their practical effectiveness.

\begin{table}[t]
    \color{black}
    \centering
    \caption{Computation time in MATLAB for Two Setups.}
    \vspace{-0.25cm}
    \label{tab:CPU_time}
    \small
    \begin{tabular}{@{}l
      S[table-format=4.2] 
      S[table-format=5.2]  
      S[table-format=5.2]  
      S[table-format=5.2]@{}} 
    \toprule
    & \multicolumn{1}{c}{Fig.~\ref{fig:PvsSE} Setup}
    & \multicolumn{1}{c}{Fig.~\ref{fig:PvsSE_Cov} Setup} \\
    \cmidrule(lr){2-2}\cmidrule(l){3-3}
    {Algorithm} & {CPU Time (ms)} & {CPU Time (ms)} \\
    \midrule
    Proposed                                  & 0.24   & 0.39    \\
    Proposed (dynamic)                        & 0.34   & 0.73    \\
    OMP~\cite{el2014spatially}                & 1.74   & 438.12 \\
    FC~\cite{sohrabi2016hybrid}               & 995.47 & 56824.39 \\ 
    PC~\cite{hu2023joint}                     & 1.23   & 24.35    \\
    \bottomrule
    \end{tabular}
\end{table}
{\color{black}To empirically verify algorithm complexity, we assess computation time in MATLAB on a workstation with i9-13900K CPU and 64GB RAM.
Table~\ref{tab:CPU_time} shows that our proposed algorithms demonstrate significant computational advantages over the baselines.
Despite the additional antenna allocation step, Proposed (dynamic) incurs a marginal increase in computation time compared to its fully-connected version, indicating that our $\CMcal{O}(N_{\sf RF}N \log N)$ partitioning algorithm adds negligible complexity to the overall framework.
}

\section{Conclusion} \label{sec:conclusion}
\vspace{-0.1cm}
We proposed the low-complexity hybrid precoding framework for MU-MIMO systems.
Our approach first identified the near-optimal RF precoder by leveraging the low-dimensional subspace property, and then optimized the digital precoder via R-WMMSE.
To deal with practical hybrid scenarios, we further extended our proposed framework in two key directions: $(i)$ dynamic-subarray architectures with adaptive antenna partitioning and $(ii)$ the channel covariance-based approach that ensures robust performance when statistical CSIT is used for designing the RF precoder.
These findings confirm that our proposed algorithms effectively provide practical hybrid precoding solutions with both improved performance and reduced complexity through revisiting the hybrid precoding design with low-dimensional subspace perspective.

\bibliographystyle{IEEEtran}
\bibliography{ref}

\end{document}